\documentclass[a4paper, 11pt]{article}
\usepackage{authblk}
\usepackage{latexsym,amsmath,amsfonts,amssymb,amsbsy,amsopn,amstext,amsthm}
\usepackage{epsfig,epstopdf,graphicx}
\usepackage[mathscr]{eucal}
\usepackage{color,multicol,makeidx}
\usepackage{bm,mathrsfs}
\usepackage{float}
\usepackage{mathrsfs}
\usepackage{tabu}
\usepackage{enumerate}
\usepackage{diagbox}

\def\eqref#1{(\ref{#1})}

\def\xra{\xrightarrow}

\DeclareMathOperator*{\argmin}{arg\,min}

\newtheorem{thm}{Theorem}
\newtheorem{lemma}{Lemma }
\newtheorem{rem}{Remark}
\newtheorem{assum}{Assumption}

\newtheorem{prop}{Proposition}
\newtheorem{exmp}{Example}
\newtheorem{defn}{Definition}

\providecommand{\keywords}[1]
{
	\small	
	\textbf{\textit{Keywords---}} #1
}

	\title{Identification of Switched Linear Systems: Persistence of Excitation and Numerical Algorithms}
	
	\author{Biqiang Mu\\
		
		Key Laboratory of Systems and Control \\
		Institute of Systems Science\\
		Academy of Mathematics and Systems Science \\
		Chinese Academy of Sciences\\
		 Beijing 100190, China \\
		(bqmu@amss.ac.cn)\\
		
		\vspace{5mm}
		Tianshi Chen\\
		\vspace{3mm}
		School of Data Science and
		Shenzhen Research Institute of Big Data\\
		The Chinese University of Hong Kong\\
		Shenzhen 518172,
		China\\
		(tschen@cuhk.edu.cn)\\
		
		\vspace{5mm}
		Changming Cheng\\
		\vspace{3mm}
		State Key Laboratory of Mechanical Systems and Vibration\\
		Shanghai Jiaotong University\\
		 Shanghai, China\\
		(ccming@sjtu.edu.cn)
		
			\vspace{5mm}
		Er-Wei Bai\\
		\vspace{3mm}
		Department of Electrical and Computer Engineering\\
		University of Iowa\\
		 Iowa City, Iowa 52242\\
		(er-wei-bai@uiowa.edu)
}

\begin{document}
	
	\maketitle
	\newpage
	\begin{abstract}
		This paper investigates two issues on identification of switched 
		linear systems: persistence of excitation and numerical algorithms. 
		The main contribution
		is a much weaker condition on the regressor to be persistently exciting that guarantees the 
		uniqueness of the
		parameter sets and also provides new insights in understanding the relation among different 
		subsystems. It is found that for uniquely determining the parameters of switched linear systems, 
		the minimum number of samples needed derived from our condition is much smaller than that 
		reported in the literature. The secondary contribution of the paper concerns the numerical algorithm.
		Though the algorithm is not new, we show that our surrogate problem, relaxed from an integer 
		optimization to a continuous minimization, has exactly the same solution as the original
		integer optimization, which is effectively solved by a block-coordinate descent algorithm. Moreover, an algorithm for handling unknown number of subsystems is proposed.
		Several numerical examples are illustrated to support theoretical analysis.
		
	\end{abstract}

\keywords{
	Switched linear systems, Persistence of excitation, Numerical algorithms.
}

	\section{Introduction}
	
	Switched linear (SL) systems  are composed of a finite number of linear subsystems but 
	operate on one of subsystems at each discrete time according to a  switching mechanism governed by a discrete decision variable that could be arbitrary or depend on the range of the regressor, and so on \cite{Paoletti2007,Lauer2019,Garulli2012}.
	SL systems might become  some other prevalent systems,  e.g., piecewise autoregressive exogenous 
	systems (PWARX), piecewise affine  systems,  hybrid linear systems, and among others 
	when the switching mechanism is specified in detail \cite{Garulli2012}. In particular, 
	the equivalence between piecewise linear systems and several classes of 
	hybrid systems is established \cite{Heemels2001}.
	Over the last two decades,  identification of SL systems has received great attention in the system identification community  since SL systems can well model many practical systems following a clear switching rule, e.g,  pulse-width-modulated driven boost converter \cite{DeKoning2003,Sun2005} and can also uniformly approximate continuous nonlinear dynamics with arbitrary accuracy \cite{Lin1992,Sontag1981,Breiman1993}. 
	
	\subsection{Problem Statement}
	This paper considers
	the SL system with a regression form described as follows \cite{Bako2011a,Lauer2018}
	\begin{align}\label{sls}
	y_k=x_k^T \theta^*_{\zeta_k^*}+\varepsilon_k,    ~~k=1,2,...,N,
	\end{align}
	where $\varepsilon_k$ is a random noise sequence,
	and $y_k\in \mathbb{R},x_k \in \mathbb{R}^n$ are 
	the measurement output and the regressor independent of the noise at index $k$.
	At each $k$, the time dependent parameter
	vector $\theta^*_{\zeta_k^*}$ chooses one from the set $\{\theta^*_1,...,\theta^*_{S}\}$ of
	cardinality $S$ representing the parameters of all subsystems. 
	The discrete finite range
	function $\zeta_k^*: k \rightarrow \{1,...,S\}$ indicates which subsystem generates
	the output at index $k$, often called the switching sequence.
	
	The goal of identifying the SL system \eqref{sls} is to estimate both the true parameter vectors $\{\theta^*_1,...,\theta^*_{S}\}$ of the $S$ subsystems and the switching sequence $\{\zeta_1^*,\cdots,\zeta_N^*\}$ for all times $k$ as accurately as possible by the data $\{x_k,y_k,k=1,\cdots,N\}$.
	
	The difficulty in identifying the SL system \eqref{sls} lies in the switching sequence $\{\zeta_k^*\}$ is unknown in advance, which has to be estimated by the data.
	Aiming at the different settings of the switching sequence, the identification algorithms for the SL systems might be different by encoding the available prior knowledge of the switching sequence.
	Therefore, the identification of the SL systems might be thought of as a preliminary estimate for other systems and  additional identification targets.
	For example, identifying the partition of the regressor space for the PWARX after having finishing the goal by first treating the system as an SL system, might be achieved using classification-based techniques \cite{Ferrari-Trecate2001} even when the estimated switching sequence is not separable \cite{Bredensteiner1999}. This strategy is adopted in the paper for 
	identifying a hybrid systems in the class of piecewise ARX systems \cite{Juloski2005}.
	
	\subsection{Existing identification algorithms}
	In the literature, most identification papers of SL systems or other related systems focus on developing identification algorithms for estimating the parameters of subsystems and the switching sequence/the partition of the regressor space.
	Some excellent survey papers \cite{Garulli2012,Paoletti2007} and book \cite{Lauer2019} on this topic are available in the literature.
	Roughly speaking, the proposed approaches can be categorized as
	
		\begin{enumerate}[1)]
			\item \textit{Optimization-based approach}. 
			This approach formulates the identification problem as a  non-convex optimization problem,
			which is solved by either  the global optimization algorithms, e.g., general branch-and-bound scheme \cite{Roll2004,Lauer2015,Lauer2018},  or the local optimization algorithms \cite{Lauer2013}.
			
			\item \textit{Regularization-based approach}. 
			This approach is to establish a loss function consisting of a prediction error term  and an additional regularization term on the parameters, where the regularization term, which can be sum-of-norms \cite{Ohlsson2011},  $\ell_2$-norm/$\ell_1$-norm \cite{Mattsson2016},  or $\ell_0$-norm/$\ell_1$-norm \cite{Bako2011a},  encodes different prior knowledge of SL systems.

			\item \textit{Classification-based approach}. 
			This approach is first to estimate  the partition or equivalently the switching sequence  by the data clustering technique
			and then to estimate the parameters of subsystems   with the estimated partition and the data \cite{Ferrari-Trecate2001,Nakada2005,Bako2011,Pillonetto2016} .

			
			\item \textit{Bayesian approach}. 
			This approach  is first to choose a prior distribution of parameters based on available prior knowledge and then to produce maximum a posteriori estimates of the parameters and the partition) to hybrid systems in PWARX \cite{Juloski2005}. 
			
			\item \textit{Bounded error approach}. 
			This approach is inspired by ideas from set-membership identification \cite{Milanese1991}, which enforces that the prediction error for all the samples is bounded by a priori quantity  \cite{Bemporad2005,Lauer2018}, where
			the priori quantity is a tuning parameter and plays a role in the balance between model accuracy and model complexity.
			
			\item \textit{Algebraic approach}. 
			The approach lifts multiple subsystems as a single but more complex linear system by applying the so-called hybrid decoupling polynomial involving the regressors and output, which is independent of the switching sequence.  Thus, a recursive identification algorithm is proposed to estimate the hybrid system parameters   in \cite{Vidal2003,Vidal2004,Vidal2008}, from which  the parameters of all the 
			subsystems are recovered  by differentiating the hybrid decoupling polynomial.
		\end{enumerate}

	\subsection{Contributions}
	
	We will study two issues in this paper including  persistence of 
	excitation and numerical algorithms. The main contribution of the paper is a weaker condition on the 
	PE of the regressor that guarantees the uniqueness of the parameter set.
	
			The persistence of excitation (PE) issue is
	a theoretic foundation for developing algorithms to recover the parameters of the subsystems and the switching sequence/the partition of the regressor space by the data.
	Despite extensive research on identification on the SL system, a little attention is paid for 
	persistence of excitation issue except for the papers based on the algebraic approach \cite{Vidal2004,Vidal2003,Vidal2008}
	and sparse optimization \cite{Bako2011a}.
	Some related discussion on the structural
	identifiability of the SL system having a state space form is provided in \cite{Petreczky2010}.
	
	For uniquely determining the parameters of the SL systems  by the data,
	the PE condition in \cite{Vidal2004,Vidal2003,Vidal2008} (Vidal's condition) implies that the minimum number of samples  has to grow exponentially with respect to both $n$ and $S$ while the PE  condition  in \cite{Bako2011a} (Bako's condition) suggests that the minimum number of samples is linear in $n$ and quadratic in $S$. 
	In the paper, we derive a new persistence of excitation condition (our condition), which derives that  the minimum sample size  is about one half of that given by Bako's condition.
	Therefore, our condition is tighter than Bako's condition and Vidal's condition.

	This paper solves the optimization problem
	\begin{subequations}\label{min2}
		\begin{align}
		&\min_{\theta_s,\xi_{s,k}} 
		\sum_{k=1}^N\sum_{s=1}^{S} 
		\xi_{s,k}(y_k-x_k^T\theta_s)^2,\label{cf}\\
		&\mbox{subject~to }\xi_{s,k} \in \{0,1\},~ \sum_{s=1}^{S} \xi_{s,k} = 1
		\end{align}
	\end{subequations}
	for seeking the parameters of all the subsystems and the switching sequence,
	where $\{\xi_{s,k}\}$ (called the membership indices) is an equivalent expression of the switching sequence $\{\zeta_1,\cdots,\zeta_N\}$.
	The problem \eqref{min2} has appeared in \cite{Lauer2013,Paoletti2007,Lauer2019} and is not new.
	This paper formulates a surrogate problem, in which a penalty term is added to the cost 
	function \eqref{cf} and the binary variables $\{\xi_{s,k}\}$ is relaxed to continuous variables $\{0\leq \xi_{s,k}\leq 1\}$.
	It is shown that the solutions of the  surrogate problem are the same as that of the original
	integer optimization \eqref{min2}. 
	Further, it is shown in the paper that
	any limit point generated by the block-coordinate descent algorithm is a stationary point of the 
	surrogate problem. When the the number $S$ of subsystems
	is unknown, an estimate for the $S$ is given by minimizing a weighted sum of data fit and model complexity.
	It is proved that the estimate converges to the true $S$ as $N\xra{}\infty$.
	
	The paper is organized as follows. 
	Section 2  aims to give a new persistence of excitation condition for guaranteeing uniqueness of
	the parameter sets.
	Numerical algorithms for seeking the parameters of all subsystems and membership indices for both known and unknown $S$ are developed in Section 3.  
	Numerical examples are provided
	in  Section 4.
	Finally, some concluding remarks are given in Section 5. 
	
	\section{Persistence of Excitation}
	\label{sec2}
	Identifiability and persistence of excitation are closely related but different concepts. 
	Conventionally, identifiability implies that any two different parameter sets 
	do not yield two models having exactly the same input-output behavior at least for some rich enough
	input signals. Further, richness of input signal is traditionally linked to persistence
	of excitation of the regressos. In the literature, it is often the case that 
	if the regressors are PE, the parameter
	set can be uniquely determined, sometime also together with a guaranteed convergence rate.   
	For the SL system, however,  the PE  is different from that of the traditional counterpart as follows:
	\begin{enumerate}[1)]
		\item it depends on not only the regressors $\{x_k,k=1,\cdots,N\}$ but also the switching sequence $\{\zeta_{k}^*,k=1,\cdots,N\}$;
		
		\item 	it has to uniquely determine not only the system parameters $\{\theta_s^*,s=1,\cdots,S\}$ but also
		the switching sequence $\{\zeta_{k}^*,k=1,\cdots,N\}$ by the data;
	\end{enumerate}
	Previously, the PE problem  is dependent on the proposed algorithms, e.g., 
	\cite{Vidal2004,Bako2011} 
	Thus, conditions to establish PE regressors are different by using different approaches. 
	This paper intends to derive conditions 
	that are not algorithm-dependent but only rely on the regressors $x_k$'s and their membership indices  $\{1,2\cdots,S\}$ . 
	The conditions derived in this paper
	are  weaker than those reported in the literature, which can deepen the understanding how the switching mechanism affects the PE property of the SL system .

	Suppose that the number $S$ of the subsystems is available.
	In the absence of noise $\varepsilon_k \equiv 0$, the PE problem of the SL system becomes whether or not the problem \eqref{min2}
	has a unique solution with respect to $\{(\theta_s,\xi_{s,k}),s=1\cdots, S,k=1,\cdots,N\}$ given the data $\{x_k,y_k,k=1,\cdots,N\}$.

	Firstly, the true system parameters $\{\theta_s^*,s=1,\cdots,S\}$ and 
	membership indices $\{\xi_{s,k}^*,s=1,\cdots, S,k=1,\cdots,N\}$
	is a solution to \eqref{min2}. 
	Secondly, let $\{(\theta_s,\xi_{s,k}),s=1\cdots, S,k=1,\cdots,N\}$ be any solution to the problem (\ref{min2}), which perfectly fits the data. 
	Thus any permutation $\{(\theta_{i_s}, \xi_{i_s,k}),s=1\cdots, S,k=1,\cdots,N)\}$ of $\{(\theta_s,\xi_{s,k}),s=1,\cdots, S,k=1,\cdots,N\}$ over the membership  index $s$
	is also a solution to (\ref{min2}), where $(i_1,i_2,\cdots,i_{S})$ is a permutation of $(1,2,\cdots,S)$.
	
	Consequently, we introduce the definition of persistence of excitation for the SL system \eqref{sls} as follows.
	
	\begin{defn}
		Let $\varepsilon_k \equiv 0$.
		We say given $n$ and $S$, the regressors
		$\{x_k,k=1,\cdots,N\}$ and the membership indices $\{\xi_{s,k}^*,s=1,\cdots, S,k=1,\cdots,N\}$ 
		are persistently  exciting (PE) for the  SL  system \eqref{sls} 
		if and only if the 
		solution of (\ref{min2}) is unique upto a permutation of $(1,2,\cdots,S)$. 
	\end{defn}
	
	Now, we aim to enter into the main part of the paper that is to derive the desired PE condition  for the  SL system \eqref{sls}.
	
	Let us denote the set
	\begin{align}
	\mathscr{C}_s^*=\{k|\xi_{s,k}^* =1,k=1,\cdots,N\}\label{cs}
	\end{align}
	for each $s\in\{1,\cdots,S\}$.
	Thus the set $\{x_k,k\in\mathscr{C}_s^*\}$ includes all the regressors associated with the parameter $\theta_s^*$.
	To investigate the PE of the regressors, some  
	necessary conditions are straightforward, e.g.,
	\begin{enumerate}[1)]
		\item $\theta^*_i\not = \theta^*_j$ for $i \not=j$;
		\item there does not exist some regressor $x_k$ with $k\in\{1,2,\cdots,N\}$ such that $x_k^T(\theta_i^*-\theta_j^*)=0$ for some $1\leq i\neq j\leq S$;
		\item 	the regressors $\{x_k,k\in\mathscr{C}_s^*\}$  for each $s\in \{1,2,\cdots,S\}$ is
		PE in the traditional identification sense for a single linear system, i.e., for each $s$,
		\begin{align}
		\label{wpe}
		\sum_{k=1}^N \xi_{s,k}^* x_kx^T_k 
		=\sum_{k\in \mathscr{C}_s^*} x_{k}x^T_{k}
		\end{align}
		is nonsingular. Otherwise, $\theta_s$'s cannot be
		uniquely determined.
	\end{enumerate}
	
	Here, it naturally raises a question: 
	whether or not the regressors  and their membership indices are PE for the SL system \eqref{sls} if  conditions 1) -- 3) above hold?
	Unfortunately, the following example shows that this conjecture is false. 
	
	\begin{figure}
		\includegraphics[scale=0.45]{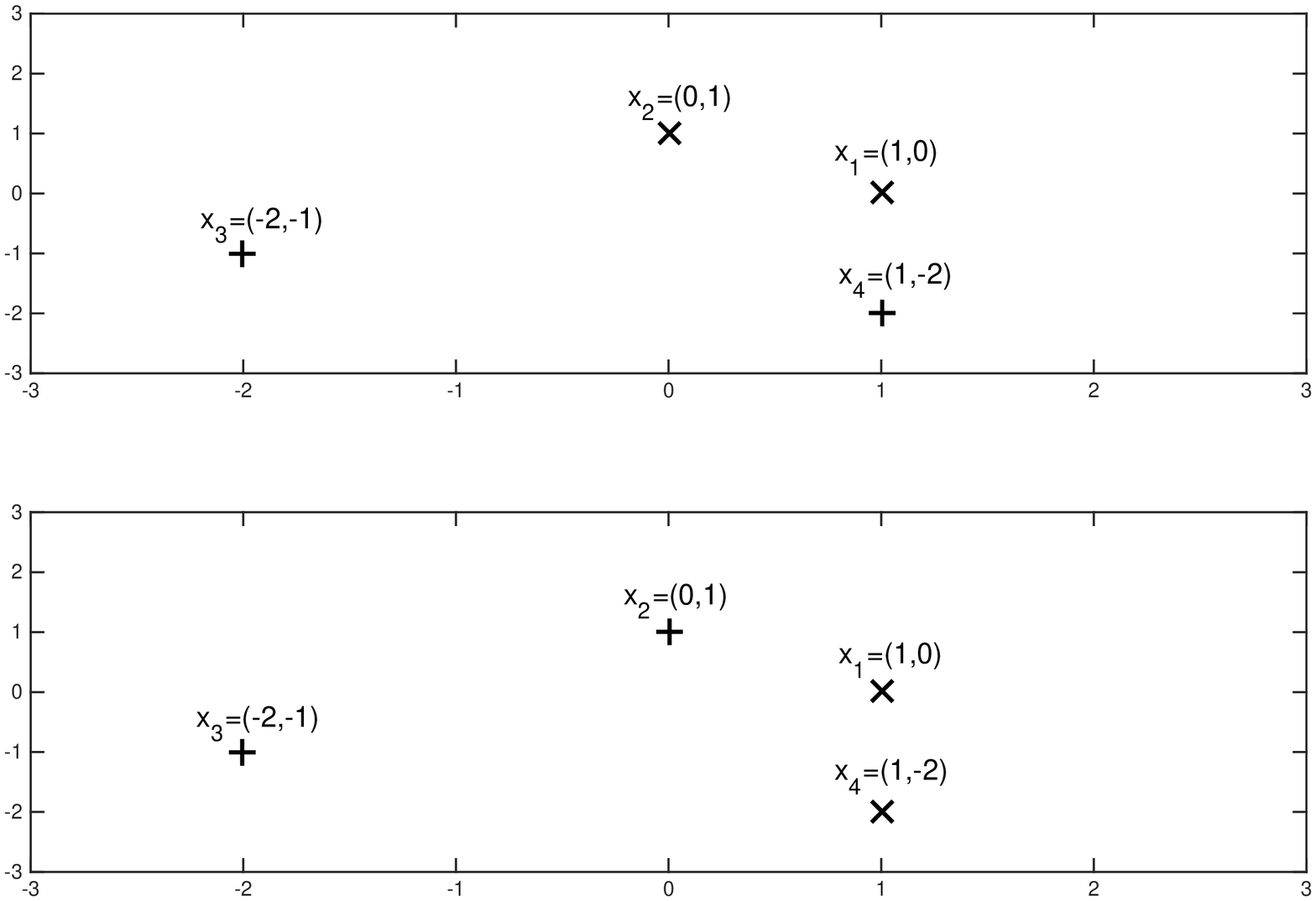}
		\caption{PE for all subsystems does not guarantees PE for the SL system}
	\end{figure}
	
	\noindent
	\begin{exmp}\label{exam1}
		\rm
		Consider the SL system having two subsystems with
		the regressors
		$
		x_{1}= \left( 1, 0 \right)^T,
		x_{2}= \left( 0, 1 \right)^T,
		x_{3}= \left( -2, -1  \right)^T,
		x_{4}= \left( 1, -2 \right)^T,
		$ 
		and the membership indices
		\begin{subequations}
			\label{ind}
			\begin{align}
			& \xi^*_{1,1}=1,~ \xi^*_{2,1}=0;~~
			\xi^*_{1,2}=1,~ \xi^*_{2,2}=0; \\
			& \xi^*_{1,3}=0,~ \xi^*_{2,3}=1; ~~
			\xi^*_{1,4}=0,~ \xi^*_{2,4}=1. 
			\end{align}
		\end{subequations}
		The corresponding parameters of the two subsystems are
		\begin{align}
		\label{pa}
		\theta^*_1 = \left( 1,1 \right)^T,~
		\theta^*_2 = \left( -2, 4 \right)^T.
		\end{align}
		It is easily verified that  regressors 
		$\{x_{1}, x_{2}\}$  regressors are PE for $\theta_1^*$, i.e.,
		$ x_{1}x^T_{1}+x_{2}x^T_{2}$ is nonsingular.
		Similarly, the regressors $\{x_{3}, x_{4}\} $  are also PE.
		The regressors $\{x_k,1\leq k \leq 4\}$ are shown in Fig. 1. 
		Now we can find that another group of parameters 
		\begin{align*}  
		\theta_1 = \left(  -0.5, 1  \right)^T,~
		\theta_2 = \left( 1,  5.5 \right)^T
		\end{align*}
		and the regressors with the membership indices
		\begin{align*}
		& \xi_{1,1}=1,~ \xi_{2,1}=0;~~ 
		\xi_{1,2}=0,~ \xi_{2,2}=1 \\
		& \xi_{1,3}=0,~ \xi_{2,3}=1; ~~
		\xi_{1,4}=1,~ \xi_{2,4}=0
		\end{align*}
		return the same outputs as the parameters \eqref{pa} and membership indices \eqref{ind} above.
		
		Thus,  there are two different solutions
		to the problem \eqref{min2}, which perfectly fit the data $\{x_k,y_k,k=1,\cdots,4\}$, and they are not a permutation of each other. 
		This means that the PE for each subsystems does not guarantee the PE for the SL   system and in fact  stronger conditions on regressors and their membership indices are needed. 
	\end{exmp}
	
	\subsection{A New PE Condition}
	To derive the desired PE conditions,  one has to further investigate the membership indices of the regressors, which are specified by the sets $\mathscr{C}_s^*,s=1,\cdots,S$.
	Let us introduce partitions of the index sets $\mathscr{C}_s^*$ defined in \eqref{cs}.
	Given an $s\in \{1,\cdots,S\}$, let $\{\mathscr{C}_s^{*(v,i)},i=1,\cdots, v\}$ with $v\in \{1,\cdots,S\}$ be a disjoint partition of $\mathscr{C}_s^*$  as follows: 
	\begin{align}
	\mathscr{C}_s^* = \bigcup_{i=1}^{v} \mathscr{C}_s^{*(v,i)}
	~\mbox{and}~
	\mathscr{C}_s^{*(v,i)}\bigcap \mathscr{C}_s^{*(v,l)}=\emptyset ~\mbox{if}~i\not=l.
	\label{pcs}
	\end{align}
	Here some $\mathscr{C}_s^{*(v,i)}$'s are allowed to be empty. 
	\begin{rem}
		Since the true membership index is unavailable for each regressor in advance, for deriving the PE condition each set $\mathscr{C}_s^*$ is partitioned into at most $S$ disjoint subsets with all possible ways subsets to enumerate the possibilities.
	\end{rem}
	
	Now, we make the following assumptions on the true parameters, regressors, and the membership indices for deriving the PE conditions.
	\begin{assum}\label{iden}
		\begin{enumerate}[1)]
			
			\item $\theta^*_i \not= \theta^*_j, i\not= j$, for $i,j \in \{1,2,...,S\}$.
			
			\item Each regressor $x_k$  satisfies $x_k\in \mathbb{R}^n\setminus Q$, where $Q=\bigcup_{ij} Q_{ij}$ and 
			$Q_{ij} = \{ x|x^T(\theta^*_i-\theta^*_j)=0, i\not= j\}$ for given $i,j\in\{1,2,\cdots,S\}$.

			\item There exists an ordered set sequence $(\mathscr{C}_{p_1}^*,\cdots,
			\mathscr{C}_{p_{S}}^*)$ in the order of $(p_1,\cdots p_{S})$ with 
			$(p_1,\cdots p_{S})$ being a permutation  of $(1,\cdots,S)$ 
			such that the following statements hold sequentially:
			\begin{enumerate}[a)]
				\item For any partition $\{\mathscr{C}_{p_1}^{*(S,i)},i=1,\cdots, S\}$ having the form  \eqref{pcs} of  $\mathscr{C}_{p_1}^*$,
				there exists some subset $\mathscr{C}_{p_1}^{*(S,i_1)}$ with $i_1\in \{1,\cdots,S\}$ such that
				$\sum_{k\in\mathscr{C}_{p_1}^{*(S,i_1)}} x_{k}x^T_{k} $ is nonsingular.
				
				\item For any partition $$\big\{\mathscr{C}_{p_s}^{*(S-s+1,i)},i \in \{1,2,\cdots,S\} 
				\setminus\{i_1,\cdots,i_{s-1}\}\big\}$$  having the form \eqref{pcs} of each 
				$\mathscr{C}_{p_s}^*$ with $s\in \{2,\cdots,S\}$,
				there exists some subset $\mathscr{C}_{p_s}^{*(S-s+1,i_s)}$ with $ i_s \in \{1,2,\cdots,S\} \setminus\{i_1,\cdots,i_{s-1}\}$ such that
				$\sum_{k\in\mathscr{C}_{p_s}^{*(S-s+1,i_s)}} x_{k}x^T_{k} $ is nonsingular.
				\item $\sum_{k\in\mathscr{C}_{p_{S}^*}} x_{k}x^T_{k} $ is nonsingular.
			\end{enumerate}
		\end{enumerate}
	\end{assum}

	The following auxiliary lemma is needed before displaying the PE condition.
	\begin{lemma}
		\label{xi}
		Consider the minimization problem (\ref{min2}). Let $\varepsilon_k \equiv 0$ and $\{(\theta_s,\xi_{s,k}),s=1,\cdots, S,k=1,\cdots,N\}$  be any solution to (\ref{min2}). Thus, under Assumption \ref{iden} there hold that
		\begin{enumerate}[1)]
			\item if $\theta_s=\theta^*_s$, $s=1,...,S$, then $\xi_{s,k}=\xi^*_{s,k}$ for all $s$ and $k$;
			
			\item if $\xi_{s,k}=\xi^*_{s,k}$ for all $s$ and $k$, then $\theta_s=\theta^*_s$ for all $s$.
		\end{enumerate}
	\end{lemma}
	
	\noindent
	{\it Proof:} For the first part, consider the prediction error $(y_k-x^T_k\theta^*_s)^2$. There are two possibilities
	for each $k$:
	\begin{enumerate}[1)]
		\item $x_k \in \mathscr{C}_s^*$ $\Longrightarrow y_k-x^T_k\theta^*_s=0$. We now claim that
		for all $j \not= s$, $(y_k-x^T_k\theta^*_j)^2>0$. If not, 
		$$
		y_k-x^T_k\theta^*_s=0=y_k-x^T_k\theta^*_j
		$$
		which implies $x_k^T(\theta^*_s-\theta^*_j)=0$ and a violation of
		the second condition of Assumption \ref{iden} occurs. The necessary condition
		$\xi_{j,k}(y_k-x^T_k\theta^*_j)^2=0$ implies $\xi_{j,k}=0$ for all $j \not =s$
		and so $\xi_{s,k}=1=\xi^*_{s,k}$. 
		
		\item $x_k \not\in \mathscr{C}_s^*$ but in $\mathscr{C}_j^*$ for some $j \not=s$ 
		implies that $y_k-x^T_k\theta^*_j=0,
		j \not=s$. By the same argument, $\xi_{j,k}=1$ and $\xi_{s,k}=0=\xi^*_{s,k}$ for
		$s \not= j$.
	\end{enumerate}
	
	For the second part,  we have
	\begingroup
	\allowdisplaybreaks
	\begin{align*}
	&\sum_{k=1}^N\sum_{s=1}^{S} \xi^*_{s,k}(y_k-x^T_k\theta_s)^2\\
	&=\sum_{j\in\mathscr{C}^*_1} (y_{j}-x^T_{j}\theta_1)^2+...
	+\sum_{j\in\mathscr{C}^*_{S}} (y_{j}-x^T_{j}\theta_{S})^2.
	\end{align*}
	\endgroup
	By the assumption that the regressors $\{x_j,j\in\mathscr{C}^*_s\}$ for each subsystem under Assumption \ref{iden} are PE, the minimization
	$\sum_{j\in\mathscr{C}^*_s} (y_{j}-x^T_{j}\theta_s)^2=0$ is achieved if and only if
	$\theta_s=\theta^*_s$ for all $s$. 
	
	This completes the proof.
	
	\begin{thm}
		\label{thmiden}
		Let $\varepsilon_k \equiv 0$ and suppose Assumption \ref{iden} holds.
		Then, 
		the regressors
		$\{x_k,k=1,\cdots,N\}$ and the membership indices $\{\xi_{s,k}^*,s=1,\cdots, S,k=1,\cdots,N\}$ 
		are PE  for the SL system  \eqref{sls}.
	\end{thm}
	{\it Proof:} 
	Let $\{(\theta_s,\xi_{s,k}),s=1,\cdots, S,k=1,\cdots,N\}$ 
	be any solution to (\ref{min2}), namely,
	$$ \sum_{k=1}^N\sum_{s=1}^{S} \xi_{s,k}(y_k-x^T_k\theta_s)^2=0.$$
	It is clear that $\theta_s\neq \theta_t, 1\leq s\neq t\leq S$. 
	Thus for proving the  theorem, it suffices to show that $\{\theta_1,\cdots,\theta_{S}\} = \{\theta^*_1,\cdots,\theta^*_{S}\}$ by using the conclusion 1) of Lemma \ref{xi}, where $\theta^*_1,\cdots,\theta^*_{S}$ are true parameters of the $S$ subsystems of the SL system \eqref{sls}.
	
	Firstly, consider the index set $\mathscr{C}_{p_{1}}^*$.
	Thus the condition that $\{(\theta_s,\xi_{s,k}),s=1,\cdots, S,k=1,\cdots,N\}$  is a solution to (\ref{min2}) derives that
	there exists some $\theta_{k_s}$ with $k_s\in\{1,2,\cdots,S\}$ such that $y_k-x^T_k\theta_{k_s}=0$ for each $k\in \mathscr{C}_{p_1}^*$.
	Thus we can divide the set $\mathscr{C}_{p_1}^*$ into $S$ subsets as follows:
	\begin{align}
	\mathscr{C}_{p_1}^{*(S,i)}= \{k\in \mathscr{C}_{p_1}^*| y_k-x^T_k\theta_{i}=0\},~i=1,\cdots,S.\label{d1}
	\end{align}
	It follows that the sets $\{\mathscr{C}_{p_1}^{*(S,i)}, i=1,\cdots,S\}$ consist of a partition of $\mathscr{C}_{p_1}^*$.
	By the third condition of Assumption \ref{iden}, 
	there exists some subset $\mathscr{C}_{p_1}^{*(S,i_1)}$ of $\mathscr{C}_{p_1}^*$ with $i_1\in \{1,2,\cdots,S\}$ such that 
	\begin{align}
	\sum_{k\in \mathscr{C}_{p_1}^{*(S,i_1)}}x_{k}x^T_{k} \label{d1i1}
	\end{align}
	is nonsingular.
	Consider the equations
	\begin{align}
	y_k=x^T_k\theta,~k\in \mathscr{C}_{p_1}^{*(S,i_1)}. \label{e1}
	\end{align}
	On one hand, $\theta^*_{p_1}$ is a solution to \eqref{e1} since $\mathscr{C}_{p_1}^{*(S,i_1)}\subset \mathscr{C}_{p_1}^*$.
	On the other hand, the non-singularity of the matrix \eqref{d1i1} implies that $\theta_{i_1}=\theta^*_{p_1}$.
	Further, we will show that there is one and only one set $\mathscr{C}_{p_1}^{*(S,i_1)}$ with $i_1\in \{1,2,\cdots,S\}$ such that 
	$
	\sum_{k\in \mathscr{C}_{p_1}^{*(S,i_1)}}x_{k}x^T_{k} 
	$
	is nonsingular.
	If this is not true, then there are $l$ sets $\{\mathscr{C}_{p_1}^{*(S,i_m)},m=1, \cdots,l\}$  with $l\in \{2,\cdots,S\}$  such that all the matrices $$
	\{\sum_{k\in \mathscr{C}_{p_1}^{*(S,i_m)}}x_{k}x^T_{k}, m=1, \cdots,l\}$$
	are nonsingular.
	Thus we can conclude that $\theta_{i_m}=\theta^*_{p_1}$ for $m=1,\cdots, l$ using the same analysis.
	However, this contradicts the fact that $\theta_i\neq \theta_j$
	if $i\neq j$.
	Therefore, we have proved that
	$\theta_{i_1}=\theta^*_{p_1}$.
	
	Secondly, consider the index set $\mathscr{C}_{p_2}^*$.
	For each $k\in\mathscr{C}_{p_2}^*$, the possible solution $\theta$ to
	$y_k=x^T_k\theta$ can not be $\theta_{i_1}$ (or $\theta^*_{p_1}$).
	Otherwise, $y_k=x^T_k\theta^*_{p_1}$.
	Also $y_k=x^T_k\theta^*_{p_2}$ due to $k\in \mathscr{C}_{p_2}^*$.
	Thus the  condition 2) of Assumption \ref{iden} is violated since $x_k^T(\theta^*_{p_1}-\theta^*_{p_2})=0$.
	Let us divide the index set $\mathscr{C}_{p_2}^*$ into $S-1$ disjoint subsets as follows:
	$$
	\mathscr{C}_{p_2}^{*(S-1,i)}= \{k\in \mathscr{C}_{p_2}^*| y_k-x^T_k\theta_{i}=0\}\label{dl}
	$$
	for $i \in\{1,2,\cdots,S\}\setminus \{i_1\}$.
	Thus, the subsets $$\big\{\mathscr{C}_{p_2}^{*(S-1,i)}, i \in\{1,2,\cdots,S\}\setminus \{i_1\}\big\}$$
	form a partition having the form \eqref{pcs} of $\mathscr{C}_{p_2}^*$.
	The third condition of Assumption \ref{iden} implies that
	there exists some set $\mathscr{C}_{p_2}^{^*(S-1,i_2)}$ with $i_2\in \{1,2,\cdots,S\}\setminus \{i_1\}$ such that 
	$
	\sum_{k\in \mathscr{C}_{p_2}^{*(S-1,i_2)}}x_{k}x^T_{k} \label{d1i2}
	$
	is nonsingular. The following procedure of proving that
	$\theta_{i_2}=\theta^*_{p_2}$ is similar to that used for the case $\mathscr{C}_{p_1}^*$ above and hence it is omitted.
	
	Finally, sequentially consider the sets $\mathscr{C}_{p_s}^*$ with $3\leq s \leq S$. 
	We can prove that  
	$\theta_{i_s}=\theta^*_{p_s}$ for each $s\in \{3,\cdots, S\}$ in 
	a similar manner. 
	
	This completes the proof.
	
	
	We now make some comments on Assumption \ref{iden}.
	\begin{itemize}
		
		\item  The first assumption is obvious, otherwise, the number  of different subsystems 
		of  the system can be reduced by merging some sets of $\{\mathscr{C}_s^*,s=1,\cdots,S\}$.
		
		\item For the second assumption, $x_k \in Q$ implies
		$y_k-x^T_k\theta^*_i=y_k-x^T_k\theta^*_j$ that makes estimation of $\xi_{s.k}$
		impossible. 
		Fortunately, the set $Q$ is measure zero in $\mathbb{R}^n$ since all the components 
		of $Q$ are a linear space of dimension $n-1<n$.
		
		\item
		The last one plays a similar role in the PE condition of traditional linear system identification for a single linear system
		but it is stronger since there are $S$ different subsystems.
		
		It is clear that the third condition of Assumption \ref{iden} reduces to the traditional PE condition 
		when $S=1$, namely, $\sum_{k=1}^N x_{k}x^T_{k} $ is nonsingular. In other words, the assumption
		is necessary for $S=1$. Its necessity is also almost obvious from Example \ref{exam1} for $S=2$ and detailed explanations are given below. Though
		we do not prove that Assumption \ref{iden} is necessary for any $S$, we guess
		the assumption is rather tight. 
		
		As shown in Example \ref{exam1}, the PE condition for each 
		subsystem does guarantee  PE for the SL system   \eqref{sls}. 
		To provide more insights, re-consider Example \ref{exam1} and recall  the parameters  $\{\theta^*_1,\theta^*_2\}$ and membership indices
		$\{\xi^*_{s,k},s=1,2,k=1,\cdots,4\}$.
		It follows that the 
		first condition of Assumption \ref{iden} holds. Further, by noting that 
		$\theta^*_1-\theta^*_2=(3,-3)^T$ and $(3,-3)x_k\neq 0$ for all $k=1,\cdots,4$, the 
		regressors $\{x_k,k=1,\cdots,4\}$ satisfy the second condition.
		Now, consider the third condition.
		We have 
		$$
		\mathscr{C}_{1}^*=\{1,2\},~~\mathscr{C}_{2}^*=\{3,4\}. 
		$$
		A possible partition of $\mathscr{C}_1^*$ is 
		$$
		\mathscr{C}_1^* = \mathscr{C}_{1}^{*(2,1)} \bigcup \mathscr{C}_{1}^{*(2,2)},~
		\mathscr{C}_{1}^{*(2,1)}=\{1\},~\mathscr{C}_{1}^{*(2,2)} =\{2\}.
		$$
		Similarly
		$$
		\mathscr{C}_2^* = \mathscr{C}_{2}^{*(2,1)} \bigcup \mathscr{C}_{2}^{*(2,2)},~
		\mathscr{C}_{2}^{*(2,1)}=\{3\},~\mathscr{C}_{2}^{*(2,2)} =\{4\}.
		$$
		All the matrices $x_1x_1^T$, $x_2x_2^T$, $x_3x^T_3$ and $x_4x^T_4$ are singular and so
		the third condition is violated.
		Now, if we can add one more regressor $x_5=(\beta,\gamma)^T$  into $\mathscr{C}_1^*$ satisfying 
		$$
		\beta\neq 0,\gamma\neq 0,\beta\neq \gamma,
		$$
		where 
		$\beta\neq 0,\gamma\neq 0$ 
		are imposed to assure that both $x_1x_1^T+x_5x_5^T$ and $x_2x_2^T+x_5x_5^T$ are 
		nonsingular, and $\beta\neq \gamma$ is required to satisfy the second condition, namely, 
		to make $x_5^T(\theta^*_1-\theta^*_2)\neq 0$.
		For the third condition,   consider the partition
		$$
		\mathscr{C}_1^* = \{1,2,5\}=\mathscr{C}_{1}^{*(2,1)} \bigcup \mathscr{C}_{1}^{*(2,2)}
		$$
		$$
		\mathscr{C}_{1}^{*(2,1)}=\{1,5\},~\mathscr{C}_{1}^{*(2,2)} =\{2\}
		$$
		or
		$$
		\mathscr{C}_{1}^{*(2,1)}=\{1\},~\mathscr{C}_{1}^{*(2,2)} =\{2,5\}
		$$
		and
		$$
		\mathscr{C}_2^* = \{3,4\}.
		$$
		Thus, each partition of $\mathscr{C}_{1}^*=\{1,2,5\}$ has a PE subset and $x_3x_3^T+x_4x_4^T$ is 
		nonsingular.
		As a result, the regressors $\{x_k,k=1,\cdots,5\}$ and the corresponding membership indices are PE for the SL system given in Example \ref{exam1}  in terms of 
		Theorem \ref{thmiden} given above.
	\end{itemize} 
	
	\subsection{In Comparison with Existing PE Conditions}
	
	The PE conditions for switched ARX systems have been studied by using an algebraic approach in  \cite{Vidal2008,Vidal2004}  and formulating an $\ell_0$ optimization problem \cite{Bako2011a}, respectively.
	Their conditions also hold for the SL system \eqref{sls} by a minor modification.
	Consequently, in the following, we compare the developed PE condition (condition 3)  of Assumption \ref{iden} with that given  in  \cite{Vidal2008,Vidal2004,Bako2011a} and see their differences.
	\begin{lemma} 
		To guarantee that the regressors and the corresponding membership indices are PE for the SL system \eqref{sls},
		the minimum numbers of required samples are
		$\tbinom{n+S}{n}-1$
		and $nS^2$, respectively, for the PE conditions given  in \cite{Vidal2004,Vidal2008} and \cite{Bako2011a},
		where $\tbinom{a}{b}$ is the number of combinations of $a$ objects taken $b$ at a time.
	\end{lemma}
	{\it Proof:}
	The PE condition derived in \cite{Vidal2008,Vidal2004} follows the equation 
	$L_Sh=0$ given as the equation (10) in \cite{Vidal2003}, where $L_S$ is the matrix constructed by the regressors and $h$ is a transform of the  parameters of the $S$ subsystems with its last element being $1$.
	The size of $L_S$ is $N\times \tbinom{n+S}{n}$ and hence the $h$ is uniquely determined by the equation $L_Sh=0$ if the dimension of the null space of $L_S$ is $1$.
	As a result, the minimum sample size $N$ is $\tbinom{n+S}{n}-1$ if $L_S$ is of full row rank.
	The minimum number of samples is $nS^2$ for the PE condition given in \cite{Bako2011a}, which is directly derived by Lemma 7 of \cite{Bako2011a}.
	\begin{thm}
		\label{thm3.3}
		Suppose that 
		\begin{enumerate}[1)]
			\item the points 1) and 2) of Assumption \ref{iden}  hold;
			\item 	each matrix
			$
			\sum_k x_{k}x_{k}^T
			$
			with the summation index $k$ over  any $n$ elements in $\mathscr{C}_{s}^*$  is nonsingular for all $1\leq s\leq S$. 
		\end{enumerate}
		Thus, to ensure that the regressors and the corresponding membership indices are PE for the SL system \eqref{sls}, the minimum number of required samples  is
		\begin{align}
		\frac{(n-1)S^2 + (n+1)S}{2}.\label{ns}
		\end{align}	
	\end{thm}
	{\it Proof:}
	We prove the result \eqref{ns} by finding a partition of the indices $\{1,\cdots,N\}$ into $S$ subsets  such that the third point of Assumption \ref{iden} holds.
	Let  the numbers $|\mathscr{C}_s^*|$ of the partition  be
	$$|\mathscr{C}_s^*|=n+(n-1)(S-s)$$ for $1\leq s \leq S$,
	where $|\cdot|$means the cardinality of  a set.
	Thus under the second point of the assumption, the ordered set sequence $(\mathscr{C}_1^*,\mathscr{C}_2^*,\cdots,\mathscr{C}_{S}^*)$ satisfies the third point of Assumption \ref{iden} and hence the regressors are PE for SL system \eqref{sls} for the given partition and the corresponding sample size is
	$$\sum_{s=1}^{S}\big(n+(n-1)(S-s)\big)=\frac{(n-1)S^2 + (n+1)S}{2}.$$	
	The similarities and differences of these PE conditions are summarized as follows:
	\begin{enumerate}[1)]
		\item The minimum numbers of required samples for ensuring the PE are $\big((n-1)S+ (n+1)\big)S/2$ (our condition), $nS^2$ (Bako's condition), and $\tbinom{n+S}{n}-1$ (Vidal's condition), respectively.
		The number of Vidal's condition  grows exponentially with both $n$ and $S$ for Vidal's  condition since 
		$
		\tbinom{n+S}{n}
		\approx
		e^{n+S} 
		\sqrt{\frac{n+S}{2\pi n S}},
		$
		where the Stirling's formula  $n! \approx \sqrt{2\pi n} (n/e)^n$ is applied, the symbol $n!$ means the factorial of an integer $n$, and $e$ is the natural constant.
		While both Bako's condition and our conditions increases linearly with $n$ and quadratically with $S$ but our condition requires less samples (around one half) than Bako's condition.
For giving a visual impression, the minimum sample sizes for the PE conditions are presented in Table \ref{tab13}, from which 
we see that when $n=S=10$, the minimum number is $184755$ for Vidal's  condition  but it is $505$ and $1000$ for our  condition and Boka's condition, respectively. 
		\item Bako's condition and our condition do not impose constraints on the switching sequence except that the numbers of samples for each subsystem is greater than an explicit and uniform lower bound.  
		While Vidal's condition imposes constraints on the switching sequence in an implicit way  as mentioned in \cite{Vidal2004}.
		\item The second point of the assumption in Theorem \ref{thm3.3} is not so impractical, e.g., it is satisfied almost surely if the regressors are randomly generated from a probability distribution.
		A similar but a little stronger condition  is assumed in Lemma 7 of \cite{Bako2011a} by  $n$-genericity index $\nu_n(X)$ of $X=[x_1^T,\cdots,x_N^T]^T$ given in Definition 1 of \cite{Bako2011a}.
		Our condition is imposed on each subset $\mathscr{C}_s^*$ while Baka's condition is assumed on $\{1,\cdots,N\}$.
	\end{enumerate}
	A straightforward insight from the Bako's  condition and our condition that guarantee 
	the PE of the  SL system is that it requires more samples greater than the dimension $n$  of the parameters for each subsystem to remove the possibility that one regressor could be assigned to   more than one subsystem, e.g., in Example \ref{exam1} the regressor $x_2=(0,1)^T$ can be assigned into either subsystem $\theta_1^*=(1,1)^T$ or subsystem $\theta_2=(1,5.5)^T$  if without adding an extra regressor.

	\begin{exmp} \rm
		In order to verify Theorem \ref{thm3.3}, we provide the following example.
		Consider a switched linear system having two subsystems with  parameters $
		\theta^*_1 = (1,1,1)^T,~\theta^*_2 = ( -2, 4,1)^T
		$
		and the regressors 
		\begin{align*}
		\left(
		\begin{array}{rrr}
		1&0&0\\
		0&1&1\\
		1&3&-1\\
		2&1&1\\
		-1&2&1\\
		-2&-1&0\\
		1&-2&-1\\
		1&-1&-2
		\end{array}
		\right)
		\end{align*}
		where $\mathscr{C}_1^*=\{1,2,3,4,5\}$ and $\mathscr{C}_2^*=\{6,7,8\}$.
		The corresponding output $y$ is produced by the system with the given regressors and parameters.
		For the considered system,  we have $N=8$, $n=3$ and $S=2$. It follows that the minimum numbers of samples for our PE condition, Bako's PE condition, and Vidal's PE condition  are $8$, $12$, and $9$, respectively. 
		Therefore,  according to either Bako's condition or Vidal's condition, the $8$ regressors and resulting membership indices are not PE,  but they are PE in terms of our condition.
		
		In the Section \ref{sec4}, we will show that the parameters $\theta^*_1,\theta^*_2$ and 
		membership indices can be exactly recovered by applying the block-coordinate descent algorithms given later.
	\end{exmp}

	
	\begin{table*}[!htp]
\tiny
		\begin{tabular}{|c|lllllll|} \hline
			\diagbox[width=3.5em]{$n$}{$S$} & $1$ & $2$ & $3$ & $4$ & $5$ & $6$ &  $7$  \\ \hline
			1	&	1/1/1	&	2/4/2	&	3/9/3	&	4/16/4	&	5/25/5	&	6/36/6	&	7/49/7	\\
			2	&	2/2/2	&	5/8/5	&	9/18/9	&	14/32/14	&	20/50/20	&	27/72/27	&	35/98/35	\\
			3	&	3/3/3	&	8/12/9	&	15/27/19	&	24/48/34	&	35/75/55	&	48/108/83	&	63/147/119	\\
			4	&	4/4/4	&	11/16/14	&	21/36/34	&	34/64/69	&	50/100/125	&	69/144/209	&	91/196/329	\\
			5	&	5/5/5	&	14/20/20	&	27/45/55	&	44/80/125	&	65/125/251	&	90/180/461	&	119/245/791	\\
			6	&	6/6/6	&	17/24/27	&	33/54/83	&	54/96/209	&	80/150/461	&	111/216/923	&	147/294/1715	\\
			7	&	7/7/7	&	20/28/35	&	39/63/119	&	64/112/329	&	95/175/791	&	132/252/1715	&	175/343/3431	\\
			8	&	8/8/8	&	23/32/44	&	45/72/164	&	74/128/494	&	110/200/1286	&	153/288/3002	&	203/392/6434	\\
			9	&	9/9/9	&	26/36/54	&	51/81/219	&	84/144/714	&	125/225/2001	&	174/324/5004	&	231/441/11439	\\
			10	&	10/10/10	&	29/40/65	&	57/90/285	&	94/160/1000	&	140/250/3002	&	195/360/8007	&	259/490/19447	\\ \hline
		\end{tabular}
		\caption{The minimum numbers of samples  for guaranteeing the PE of the regressors for the SL system \eqref{sls} with  $1\leq n \leq 10$ and $1\leq S\leq 7$  (for each $n$ and $S$ the sequential numbers separated by the symbol $``/"$  represent the minimum numbers required by our  condition, Bako's condition,  and Vidal's condition, respectively).}
		\label{tab13}
	\end{table*}

	\section{Numerical Algorithms}
	\label{sec3}
	This section aims to provide a heuristic algorithm for solving the optimization 
	problem \eqref{min2}, where $y_k$ can be either the noise-free output or noisy output.  We first use a block-coordinate descent algorithm for seeking the parameters and membership indices of the SL system \eqref{sls} with a known $S$, i.e., solving the optimization problem \eqref{min2}, and then consider the case for unknown $S$.
	
	\subsection{A block-coordinate descent algorithm}
	\label{sec31}
	It is clear that the minimization (\ref{min2}) is a mixed-integer optimization
	problem, which is NP-hard. 
	Our idea is to relax $\xi_{s,k} \in \{0,1\}$ to $\xi_{s,k}\in [0,1]$
	but to add a penalty term on the membership indices for each $k$,
	\begin{subequations}\label{min4}
		\begin{align}
		\min_{\theta_s,\xi_{s,k}} &
		\sum_{k=1}^N 
		\left\{\sum_{s=1}^{S} \xi_{s,k}(y_k-x_k^T\theta_s)^2+
		\Big(1 -  \sum_{s=1}^{S} \xi_{s,k}^2\Big)\right\}\\
		&\mbox{subject~to }0\leq \xi_{s,k} \leq 1,~ \sum_{s=1}^{S} \xi_{s,k} = 1.
		\label{cons}
		\end{align}
	\end{subequations}
	We first show that the minimization problem (\ref{min4}) shares the same solutions with the minimization problem (\ref{min2}).
	\begin{lemma}
		\label{lemma32}
		Consider the minimization problem (\ref{min4}).
		Let $\{\xi'_{s,k},s=1,\cdots,S,k=1,\cdots,N\}$ be a solution of (\ref{min4}) for any fixed $\{\theta_s,s=1,\cdots,S\}$.
		Then, $\xi'_{s,k} \in \{0,1\}$ and $\sum_{s=1}^{S} \xi'_{s,k} = 1$ for each $k$.
	\end{lemma}
	
	\noindent
	{\it Proof:} 
	The conclusion is proved for each $k$ with $k=1,\cdots,  N$.
	For each $k$, the cost function of \eqref{min4} is 
	\begingroup
	\allowdisplaybreaks
	\begin{subequations}
		\begin{align}
		\label{eq1}
		&\sum_{s=1}^{S}  \xi_{s,k}(y_k-x_k^T\theta_s)^2+
		\Big(1 -  \sum_{s=1}^{S} \xi_{s,k}^2\Big)\\
		&\geq \sum_{s=1}^{S}  \xi_{s,k}\min_{1\leq s \leq S} (y_k-x_k^T\theta_s)^2
		+
		\Big(1 -  \sum_{s=1}^{S} \xi_{s,k}^2\Big)\\
		&=\min_{1\leq s \leq S} (y_k-x_k^T\theta_s)^2
		+
		\Big(1 -  \sum_{s=1}^{S} \xi_{s,k}^2\Big),
		\end{align}
	\end{subequations}
	\endgroup
	where the first equation and last equation have applied the constraint \eqref{cons}.
	Thus, we see that the cost function \eqref{eq1} achieves its minimum under the constraint \eqref
	{cons} if and only if one of $\xi_{s,k}$ is $1$ and the rest is zero.
	
	This completes the proof. 
	\begin{prop}
		\label{prop3}
		The solutions to (\ref{min2}) are the same as the solutions to  (\ref{min4}).
	\end{prop}
	
	{\it Proof:}
	Let $\{(\theta'_s,\xi'_{s,k}),s=1,\cdots, S,k=1,\cdots,N\}$ be any solution to (\ref{min4}) and let $\{(\theta_s^*,\xi_{s,k}^*),s=1,\cdots, S,k=1,\cdots,N\}$  any solution to (\ref{min2}).
	Thus there holds 
	\begingroup
	\allowdisplaybreaks
	\begin{align*}
	&\hspace{3mm}\min_{\theta_s,\xi_{s,k}}
	\sum_{k=1}^N 
	\left\{\sum_{s=1}^{S} \xi_{s,k}(y_k-x_k^T\theta_s)^2+
	\Big(1 -  \sum_{s=1}^{S} \xi_{s,k}^2\Big)\right\},\\
	&\hspace{15mm}\mbox{subject~to }0\leq \xi_{s,k} \leq 1,~ \sum_{s=1}^{S} \xi_{s,k} = 1\\
	&= \min_{\xi_{s,k}}
	\sum_{k=1}^N  \left\{\sum_{s=1}^{S} \xi_{s,k}(y_k-x_k^T\theta'_s)^2+
	\Big(1 -  \sum_{s=1}^{S} \xi_{s,k}^2\Big)\right\},\\
	&\hspace{15mm} \mbox{ subject to }\xi_{s,k} \in \{0,1\},~ \sum_{s=1}^{S} \xi_{s,k} = 1\\
	&= \min_{\xi_{s,k}}
	\sum_{k=1}^N \sum_{s=1}^{S} \xi_{s,k}(y_k-x_k^T\theta'_s)^2, 
	\mbox{ subject to } \xi_{s,k} \in \{0,1\},~ \sum_{s=1}^{S} \xi_{s,k} = 1\\
	&\geq \min_{\theta_s, \xi_{s,k}}
	\sum_{k=1}^N \sum_{s=1}^{S} \xi_{s,k}(y_k-x_k^T\theta_s)^2,\mbox{ subject to } \xi_{s,k} \in \{0,1\},~ \sum_{s=1}^{S} \xi_{s,k} = 1,
	\end{align*}
	\endgroup
	where the first equality holds by Lemma \ref{lemma32} 
	and the second equality follows from  the constraints, i.e., the penalty
	term is zero.
	
	Similarly,  we have
	\begingroup
	\allowdisplaybreaks
	\begin{align*}
	&\hspace{3mm}\min_{\theta_s, \xi_{s,k}}
	\sum_{k=1}^N \sum_{s=1}^{S} \xi_{s,k}(y_k-x_k^T\theta_s)^2, \mbox{ subject to } \xi_{s,k} \in \{0,1\},~ \sum_{s=1}^{S} \xi_{s,k} = 1\\
	&= \min_{\xi_{s,k}}
	\sum_{k=1}^N  \left\{\sum_{s=1}^{S} \xi_{s,k}(y_k-x_k^T\theta_s^*)^2\right\}, \mbox{ subject to } \xi_{s,k} \in \{0,1\},~ \sum_{s=1}^{S} \xi_{s,k} = 1\\
	&= \min_{\xi_{s,k}}
	\sum_{k=1}^N  \left\{\sum_{s=1}^{S} \xi_{s,k}(y_k-x_k^T\theta_s^*)^2+
	\Big(1 -  \sum_{s=1}^{S} \xi_{s,k}^2\Big)\right\}\\
	&\hspace{8mm}\mbox{ subject to } 0\leq \xi_{s,k} \leq 1,~ \sum_{s=1}^{S} \xi_{s,k} = 1\\
	&\geq \min_{\theta_s,\xi_{s,k}}
	\sum_{k=1}^N  \left\{\sum_{s=1}^{S} \xi_{s,k}(y_k-x_k^T\theta_s)^2+
	\Big(1 -  \sum_{s=1}^{S} \xi_{s,k}^2\Big)\right\}\\
	&\hspace{8mm}\mbox{ subject to } 0\leq \xi_{s,k} \leq 1,~ \sum_{s=1}^{S} \xi_{s,k} = 1.
	\end{align*}
	\endgroup
	
	This finishes the proof.
	
	The idea that relaxing the binary values $\xi_{s,k}\in\{0,1\}$ to continuous variables $\xi_{s,k}\in [0,1]$ is not new, e.g., see \cite{Munz2005}, but here we add  penalty terms to the cost function that did not appear in \cite{Munz2005}.
	Therefore, our conclusion of Proposition \ref{prop3} is stronger than that given in Lemma 1 of \cite{Munz2005}, which says that 
	``The global optimal value of problem (4) coincides with the global optimal value of the relaxed problem (5). The set of optimal solutions of problem (4) is a subset of the set of optimal solutions of problem (5)."
	The problems (4) and (5) in \cite{Munz2005} can be understood as the optimization problem \eqref{min2} and the problem \eqref{min4} without the penalty terms in this paper, respectively.
	
	Now, we use the block-coordinate descent algorithms \cite[Section 3.7]{Bertsekas2016}; \cite[Section 6.1.3]{Lauer2019}   to solve the problem \eqref{min4}, which is not new for estimating the SL systems \cite{Lauer2013} but works well in practice, especially for small $n$ and $S$.
	The  algorithm  iterates in the following way:
	
	Step 0: Initialize  $\{\xi_{s,k},s=1,\cdots, S, k=1,\cdots,N\}$ such that  $\xi_{s,k}\in \{0,1\}$ and $\sum_{s=1}^{S} \xi_{s,k} = 1$ for each
	$s=1,2,...,S$ and $k=1,2,...,N$, which can be produced in a random way or any clustering technique.
	
	Step 1: Given $\widehat{\xi}_{s,k}$'s, the parameter $\theta_s$ is determined by 
	\begingroup
	\allowdisplaybreaks
	\begin{subequations}
		\label{alg1}
		\begin{align}
		\widehat{\theta}_s&= \argmin_{\theta_s} \sum_{k=1}^N \widehat{\xi}_{s,k}(y_k-x^T_k\theta_s)^2\\
		&= \Big(\sum_{k\in \mathscr{B}_s} x_{k}x^T_{k}\Big)^{-1}
		\sum_{k\in \mathscr{B}_s} x_{k}y_{k},
		\end{align}
	\end{subequations}
	\endgroup
	where $\mathscr{B}_s=\{k|\widehat{\xi}_{s,k}=1,k=1,\cdots,N\}$ for $1\leq s \leq S$.
	
	
	
	\noindent
	Step 2: For $\widehat{\theta}_s$, $s=1,2,...,S$ produced from Step 1, $\xi_{s,k}$ are solved by
	\begin{align*}
	\widehat{\xi}_{s,k}=\argmin_{\xi_{s,k}}&
	\sum_{k=1}^N\left\{\sum_{s=1}^{S} 
	\xi_{s,k}(y_k-x_k^T\widehat{\theta}_s)^2+ 
	\Big(1 -  \sum_{s=1}^{S} \xi_{s,k}^2\Big)\right\}\\
	\mbox{ subject to } &0\leq \xi_{s,k}\leq 1, \sum_{s=1}^{S} \xi_{s,k} = 1.
	\end{align*}
	Actually, the solutions $\widehat{\xi}_{s,k}$s have a closed form. That is for each $k$, 
	\begin{align}
	\label{alg2}
	\widehat{\xi}_{s,k}=\left\{
	\begin{array}{cc}
	1&~if~s=\argmin\limits_{s\in\{1,2,\cdots,S\}} (y_k-x_k^T\widehat{\theta}_s)^2\\
	0&otherwise
	\end{array}\right.
	.
	\end{align}
	
	Step 3: Go back to Step 1 unless some stop criteria is met.
	\begin{rem}
		The algorithm \eqref{alg1}--\eqref{alg2} is similar to Algorithm 1 of \cite{Lauer2013} except for the alternating order at each iteration.
	\end{rem}
	Based on the convergence result given in \cite[Corollary 2]{Grippo2000} and \cite[Chapter 8]{Bard1998}, 
	we have the following conclusion on the alternating algorithm.
	\begin{prop}\label{th1}
		Consider the optimization problem (\ref{min4}) and the block-coordinate descent algorithm given in \eqref{alg1}--\eqref{alg2}.
		Then, every limit point of the sequence generated by \eqref{alg1}--\eqref{alg2} is a stationary point of (\ref{min4}).
	\end{prop}
	The computational complexity of the block-coordinate descent algorithm is linear in the sample size $N$ and quadratic in either the dimension of the parameters $n$ or the number of the subsystems $S$.
	This means that the block-coordinate descent algorithm is computationally feasible even for moderately large $n$ and $S$.

	%

	%
	%
	%
	
	\subsection{An algorithm for handling unknown $S$} 
	
	We now turn our attention to the case that $S$ is unknown. 
	If  an upper bound $\bar{S} \geq S$ is available, then
	we estimate the true $S$ by 
	\begin{align}
	\label{estS}
	\widehat{S}_N=\argmin_{1\leq S' \leq \bar{S}} \frac{1}{N} \sum_{k=1}^N\sum_{s=1}^{S'} 
	\widehat{\xi}_{s,k}(y_k-x_k^T\widehat{\theta}_s)^2+\lambda_N S'
	\end{align}
	where $\lambda_N>0$ is a tuning parameter and $\{(\widehat{\theta}_s,\widehat{\xi}_{s,k}),s=1,\cdots,S',k=1,\cdots,N\}$ is the solution to the problem
	\begin{subequations}
		\label{estS'}
		\begin{align}
		\min_{\theta_s,\xi_{s,k}} &
		\sum_{k=1}^N 
		\left\{\sum_{s=1}^{S'} \xi_{s,k}(y_k-x_k^T\theta_s)^2+
		\Big(1 -  \sum_{s=1}^{S'} \xi_{s,k}^2\Big)\right\}\\
		&\mbox{subject~to }0\leq \xi_{s,k} \leq 1,~ \sum_{s=1}^{S'} \xi_{s,k} = 1.
		\end{align}
	\end{subequations}
	
	\begin{assum}
		\label{ass4}
		\begin{enumerate}[1)]
			\item 
			Let $\{\theta_1,\cdots,\theta_{S'}\}$ be a set consisting 
			of any $S'$ parameter vectors of length $n$ for $0<S'<S$. 
			Thus
			there exists a sequence  $\{\delta_k,1\leq k \leq N\}$ such that  for each $k$ and $\beta_k$ arbitrarily taking from $\{\theta_1,\cdots,\theta_{S'}\}$
			\begin{align*}
			y_k= x_k^T \beta_k +\delta_k+\varepsilon_k
			\end{align*}
			and 
			\begin{align*}
			&\frac{1}{N} \sum_{k=1}^N \delta_k^2 \geq \epsilon  >0,
			\end{align*}
			where $\epsilon$ is a positive constant.
			
			\item The noises $\{\varepsilon_k,~k=1,...,N\}$ are a sequence of iid random variables with zero mean and finite variance $\sigma^2$.
			
			\item The upper bound $\bar{S} \geq S$ is available.
			
			\item $\lambda_N>0$ so that $\lambda_N \rightarrow 0$
			and $N\lambda_N \rightarrow \infty$ as $N \rightarrow \infty$. 
		\end{enumerate}
	\end{assum}
	\begin{rem}\rm 
		Suppose the data $\{x_k,y_k,1\leq k \leq N\}$ of the SL system \eqref{sls} is indeed generated by $S$ subsystems.
		An intuitive explanation of 1) of Assumption \ref{ass4} is that the fitting error for the data introduces an additional bias term $\delta_k$ besides the noise no matter how we choose the parameters of subsystems and the membership indices if the number of subsystems for the fitting is strictly less than $S$.
	\end{rem}
	\begin{exmp}\rm
		Now, we use an example to show that the 1) of Assumption \ref{ass4} is reasonable. Consider an  SL system \eqref{sls} with $S=2$ and
		$\xi_{1,k}^*=1$ for $1\leq k \leq N_1$ and $\xi_{2,k}^*=1$ for $N_1+1\leq k \leq N$.
		If the SL system is indeed generated in the way \eqref{sls}, namely,
		\begingroup
		\allowdisplaybreaks
		\begin{subequations}
			\label{exsls}
			\begin{align}
			&y_k= x_k^T \theta^*_1 +\varepsilon_k, ~for~ 1\leq k \leq N_1\\
			&y_k= x_k^T \theta^*_2 +\varepsilon_k, ~for~ N_1+1\leq k \leq N,
			\end{align}
		\end{subequations}
		\endgroup
		where $\theta^*_1\neq \theta^*_2$.
		It follows  that
		\begin{align*}
		&y_k= x_k^T \beta_1 + \underbrace{x_k^T (\theta^*_1 - \beta_1 )}_{\delta_k} +\varepsilon_k, ~for~ 1\leq k \leq N_1,\\
		&y_k= x_k^T \beta_1 + \underbrace{x_k^T (\theta^*_2 - \beta_1 )}_{\delta_k} +\varepsilon_k, ~for~ N_1+1\leq k \leq N
		\end{align*}
		if using only one subsystem (one parameter $\beta_1$) to fit the system \eqref{exsls}.
		We see that 
		\begin{align*}
		&\frac{1}{N} \sum_{k=1}^N \delta_k^2
		=
		\frac{1}{N}  (\theta^*_1 - \beta_1 )^T \left(\sum_{k=1}^{N_1} x_kx_k^T\right) (\theta^*_1 - \beta_1 )\\
		&+\frac{1}{N}
		(\theta^*_2 - \beta_1 )^T \left(\sum_{k=N_1+1}^{N} x_kx_k^T\right) (\theta^*_2 - \beta_1 )
		\geq \epsilon  >0
		\end{align*}
		holds no matter what value of $\theta_1$ we choose as long as  the minimum eigenvalues of
		$\frac{1}{N}\left(\sum_{k=1}^{N_1} x_kx_k^T\right)$ and $\frac{1}{N}\left(\sum_{k=N_1+1}^{N} x_kx_k^T\right)$
		have a lower bound, which is easy to be satisfied.
	\end{exmp}
	
	\begin{thm}
		\label{thm3}
		Suppose that Assumption \ref{ass4} holds. Then,
		in probability as $N\rightarrow \infty$,  there holds that
		$
		\widehat{S}_N\xrightarrow{}{}S.
		$
	\end{thm}
	{\it Proof:} 
	For $S' \geq S$, the cost function of \eqref{estS} is asymptotically equal to
	$\sigma^2+\lambda_NS'+O_p(\frac{1}{N})$ for large $N$, where $\sigma^2$ is the variance of the noise. 
	Since $N\lambda_N \rightarrow \infty$,
	the term $O_p(\frac{1}{N})$ is negligible compared to $\lambda_N$. Thus for any $S_1 \geq S_2 \geq S$
	and large enough $N$,
	we have 
	\begin{align*}
	\sigma^2+\lambda_NS_1 +O_p\Big(\frac{1}{N}\Big)
	&\geq 
	\sigma^2+\lambda_NS_2+O_p\Big(\frac{1}{N}\Big) \\
	&\geq \sigma^2+\lambda_NS+O_p\Big(\frac{1}{N}\Big).
	\end{align*}
	Consequently, among all $S' \geq S$, the minimum of the cost function of \eqref{estS} is asymptotically achieved at $S'=S$ for large $N$. 
	
	Now consider the case $1\leq S' <S$.
	the cost function of \eqref{estS} is asymptotically equal to
	$\sigma^2+\frac{1}{N}\sum_{k=1}^N \Delta^2_k+\lambda_NS'+O_p\Big(\frac{1}{N}\Big)$, which derives that for large
	$N$, there holds that for all $1\leq S' < S$
	\begin{align*}
	\sigma^2
	+\frac{1}{N}\sum_{k=1}^N \delta^2_k
	&+\lambda_NS' +O_p\Big(\frac{1}{N}\Big)\\
	&>
	\sigma^2+\epsilon +\lambda_NS'+O_p\Big(\frac{1}{N}\Big)\\
	&> \sigma^2+\lambda_NS+O_p\Big(\frac{1}{N}\Big),
	\end{align*}
	where $\epsilon>\lambda_N(S-S')$ holds asymptotically since $\lambda_N \rightarrow 0$.
	
	As a result, combining the case $S' \geq S$ with the case $1\leq S' < S$, the cost function of \eqref{estS} takes its minimum asymptotically at $S'=S$ for large $N$.
	
	This completes the proof.

	\section{Numerical Simulations}
	\label{sec4}
	This section provides several examples to show the performance of the proposed numerical algorithms in Section \ref{sec3}.
	
		\noindent
	\textbf{Example 2'.}
	The proposed block-coordinate descent algorithm is applied to Example 2 in Section \ref{sec2} and the results are
	given in the Table \ref{tab111}. We see that the proposed algorithm almost finds 
	the true parameters and membership indices after just 3 iterations.
	This shows that the algorithm converges very quickly if it converges.
	On the other hand, if we remove the first regressor $[1,0,0]^T$ and the 
	corresponding output, then we have  $7$ samples (less than $8$) that are  not PE by our PE condition. 
	We  rerun the same algorithm for the $7$ samples and the result shown 
	in the Table \ref{tab12}. It is found  that another group of parameters and membership indices also 
	perfectly fit the $7$ samples.
	This  indicates that our PE condition is tight.
	\begin{table*}[!htp]
		\scriptsize
		\begin{tabular}{clll} \hline
			Iters& Estimates  & &\\ \hline
					0 	& 	$\widehat{\zeta}=[1,1,2,1,1,1,1,2]$ (Random initial values for the switching sequence)\\ \hline
			1  &    $\widehat{\theta}_1=[-0.3636,1.7500,2.1250]$ and $\widehat{\theta}_2=[-1.0606,2.1515,2.3939]$, \\
			&$\widehat{\zeta}=[1,1,2,1,1,2,2,2]$,
			~~$obj = 32.5250$ \\  \hline
			2& $\widehat{\theta}_1=[1.000,1.000,1.000]$ and $\widehat{\theta}_2=[-1.6327,2.6531,2.4694]$,\\
			&	$\widehat{\zeta}=[1,1,1,1,1,2,2,2]$, ~~$obj =4.4082$\\  \hline
			3& $\widehat{\theta}_1=[1.000,1.000,1.000]$ and $\widehat{\theta}_2=[-2.000,4.000,1.000]$,\\	&$\widehat{\zeta}=[1,1,1,1,1,2,2,2]$, ~~$obj=7.8541\times 10^{-29}$\\
			\hline
			4& $\widehat{\theta}_1=[1.000,1.000,1.000]$ and $\widehat{\theta}_2=[-2.000,4.000,1.000]$, \\
			&	$\widehat{\zeta}=[1,1,1,1,1,2,2,2]$, ~~ $obj=7.8541\times 10^{-29}$\\
			\hline
		\end{tabular}
		\caption{Iterative results of the block-coordinate descent  algorithm developed in Section \ref{sec3} using  8 regressors and corresponding outputs, where
			$\widehat{\zeta}$ is an estimate for the the switching sequence at each  iteration 
			and the {\it obj} is the cost function value of the problem \eqref{min2} corresponding to the estimated parameters and switching sequence.}
		\label{tab111}
	\end{table*}

	\begin{table*}[!htp]
		\centering{\scriptsize
			\begin{tabular}{clll} \hline
				Iters& Estimates  &&\\ \hline
							0 	&	$\widehat{\zeta}=[1,1,1,2,1,1,2]$ (Random initial values for the switching sequence)\\ \hline
				1  &    $\widehat{\theta}_1=[-1.2642,2.5283,2.8679]$ and $\widehat{\theta}_2=[-0.9091,-1.6364,4.3636]$,\\
				&	$\widehat{\zeta}=[2,1,1,2,1,1,1,2]$, 
				~~$obj=5.2965$ \\\hline
				2& $\widehat{\theta}_1=[-1.4000,2.8000,4.000]$ and $\widehat{\theta}_2=[-2.0000,-2.0000,4.0000]$,\\
				&	$\widehat{\zeta}=[2,1,1,2,1,1,1,2]$,
				~~$obj=4.2796\times 10^{-29}$\\  \hline
				3& $\widehat{\theta}_1=[-1.4000,2.8000,4.000]$ and $\widehat{\theta}_2=[-2.0000,-2.0000,4.0000]$,\\
				&	$\widehat{\zeta}=[2,1,1,2,1,1,1,2]$,
				~~ $obj=4.2796\times 10^{-29}$\\
				\hline
		\end{tabular}}
		\caption{Iterative results of the algorithm developed in Section \ref{sec3} using 7 regressors and corresponding outputs. 
			The notations and meanings here are the same as that in Table \ref{tab111}.}
		\label{tab12}
	\end{table*}
	
\begin{figure}[!h]
		\includegraphics[scale=0.5]{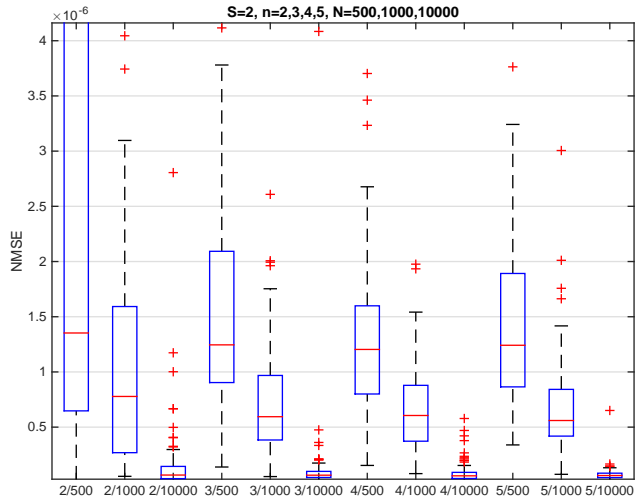}
		\caption{The boxplot of the NMSE for all the cases when $S=2$, where the symbol $2/500$ corresponds to the case $n=2$ and $N=500$ and the meanings of others are similar.}
		\label{fig41}
	\end{figure}

\begin{figure}
		\includegraphics[scale=0.5]{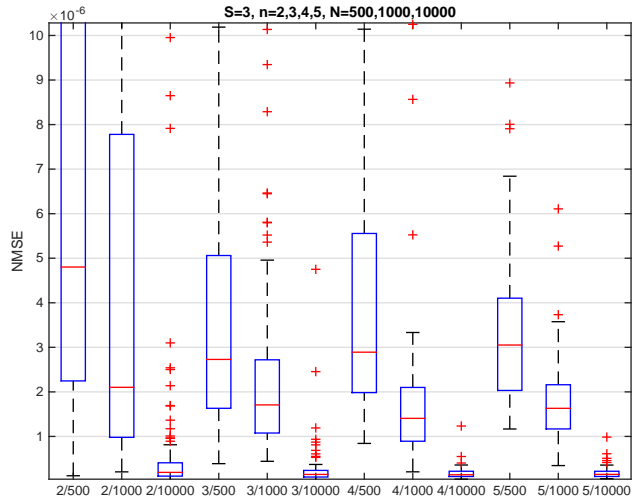}
		\caption{The boxplot of the NMSE for all the cases when $S=3$.}
		\label{fig42}
	\end{figure}

\textbf{Example 3.}
		The settings of this example is the same as that given in Section 5.1 of \cite{Lauer2018}. 
		We simulate the SL system \eqref{sls} with the cases $n=2,3,4,5$, $S=2,3$, and $N=500,1000,10000$ in the following way: 1) each element of parameters and regressors are uniformly drawn from the interval $[-5,5]$;  the switching sequence $\zeta_{k}$ is uniformly drawn from $\{1,\cdots,S\}$;
		3) the noise is a sequence of iid Gaussian random variables with zero mean and standard deviation $\sigma= 0.1$. We repeat each case 100 times. 
		The initialization of the membership indices is randomly generated from a uniform distribution.
		The performance of the parameter estimation is evaluated by the normalized parametric mean squared error, ${\rm NMSE} =\sum_{s=1}^S 
		\|\widehat{\theta}_s-\theta_s^*\|^2/\|\theta_s^*\|^2$, where $\widehat{\theta}_s$ is the estimate for the $s$-th subsystem.
		The classification error rate (CE) is defined as  the fraction of data points for which the switching sequence $\{\xi_k\}$ is incorrectly estimated.
		The number out of 100 repetitions finding the true parameters of subsystems for each case is denote by NRFTP, where one repetition is thought of  finding the true parameters if the NMSE is less than $10^{-4}$.
		The computing is implemented in a  Matlab platform running on a laptop with a 2.2 GHz i7-dual core processor.
		Average and standard deviation of the computing time, the NMSE and the CE as well as NRFTP for  the SL system with the cases given above are presented in Table \ref{tab5}. 
		Also, the boxplots of the NMSE are given in Figures \ref{fig41} and \ref{fig42}.
		It can seen from the simulation results that 
		\begin{enumerate}[1)]
			\item the computing time of the block-coordinate descent algorithm is scalable with respect to the sample size, the dimensions, and the number of subsystems.
			
			\item although the algorithm just converges to a stationary point, the simulation results have shown that it finds the true parameters with a very high probability for the considered cases.
		\end{enumerate}
	
		\begin{table}[!ht]
			\centering
			\caption{Average and standard deviation of the computing time, the NMSE and the classification error rate (CE) as well as the number out of 100 repetitions finding the true parameters of subsystems (NRFTP) for switching linear systems with subsystems $S=2,3$, dimensions $n=2,3,4,5$, sample sizes $N=500,1000,10000$.}
			\vspace{1.5ex}
			\begin{tabu} to 1.1
				\textwidth{X[.1,l]X[.2,l]X[0.8,c]X[0.7,c]X[0.9,c]X[0.25,l]}
				\hline
				$n$ & $N$&  \mbox{Time (s)}    &   CE(\%)  &     NMSE &   NRFGS  \\  \hline
				&&&$S=2$&&\\ \hline
				2& 500& 	0.0012 	$\pm$	0.0002 	&	0.942	$\pm$	3.9629 	&	0.0203 	$\pm$	0.2028 	&	2 	\\
				&1000&	0.0013 	$\pm$	0.0003 	&	1.071	$\pm$	4.6541 	&	0.0385 	$\pm$	0.3847 	&	2 	\\
				&10000&	0.0063 	$\pm$	0.0019 	&	0.4843	$\pm$	0.5262 	&	2.20E-07	$\pm$	7.38E-07	&	0 	\\
				3& 500& 	0.0013 	$\pm$	0.0002 	&	0.772	$\pm$	3.8358 	&	0.0028 	$\pm$	0.0277 	&	1 	\\
				&1000&	0.0015 	$\pm$	0.0003 	&	0.409	$\pm$	0.2644 	&	8.40E-07	$\pm$	1.15E-06	&	0 	\\
				&10000&	0.0077 	$\pm$	0.0029 	&	0.3794	$\pm$	0.3004 	&	1.22E-07	$\pm$	4.07E-07	&	0 	\\
				4& 500& 	0.0014 	$\pm$	0.0002 	&	0.274	$\pm$	0.2908 	&	1.42E-06	$\pm$	1.16E-06	&	0 	\\
				&1000&	0.0016 	$\pm$	0.0004 	&	0.286	$\pm$	0.2020 	&	6.98E-07	$\pm$	5.29E-07	&	0 	\\
				&10000&	0.0081 	$\pm$	0.0028 	&	0.2989	$\pm$	0.1785 	&	8.71E-08	$\pm$	9.41E-08	&	0 	\\
				5& 500& 	0.0016 	$\pm$	0.0014 	&	0.244	$\pm$	0.2568 	&	1.43E-06	$\pm$	7.86E-07	&	0 	\\
				&1000&	0.0018 	$\pm$	0.0004 	&	1.199	$\pm$	6.6664 	&	0.0294 	$\pm$	0.2106 	&	2 	\\
				&10000&	0.0087 	$\pm$	0.0028 	&	0.2498	$\pm$	0.1052 	&	7.22E-08	$\pm$	6.59E-08	&	0 	\\ \hline
				&&&$S=3$&&\\ \hline
				2& 500& 	0.0019 	$\pm$	0.0006 	&	6.95	$\pm$	15.2403 	&	1.4252 	$\pm$	9.8862 	&	15 	\\
				&1000&	0.0022 	$\pm$	0.0007 	&	5.275	$\pm$	14.2459 	&	0.2893 	$\pm$	1.1636 	&	10 	\\
				&10000&	0.0120 	$\pm$	0.0064 	&	2.083	$\pm$	7.4784 	&	0.0811 	$\pm$	0.5846 	&	2 	\\
				3& 500& 	0.0021 	$\pm$	0.0007 	&	1.96	$\pm$	6.5502 	&	0.0323 	$\pm$	0.2117 	&	3 	\\
				&1000&	0.0025 	$\pm$	0.0008 	&	1.713	$\pm$	6.2855 	&	0.0805 	$\pm$	0.5895 	&	2 	\\
				&10000&	0.0127 	$\pm$	0.0043 	&	0.7331	$\pm$	0.3086 	&	2.65E-07	$\pm$	5.43E-07	&	0 	\\
				4& 500& 	0.0023 	$\pm$	0.0008 	&	3.306	$\pm$	11.3625 	&	0.0885 	$\pm$	0.3758 	&	6 	\\
				&1000&	0.0027 	$\pm$	0.0008 	&	2.137	$\pm$	7.9008 	&	0.0523 	$\pm$	0.2886 	&	4 	\\
				&10000&	0.0129 	$\pm$	0.0042 	&	0.5706	$\pm$	0.2002 	&	1.77E-07	$\pm$	1.39E-07	&	0 	\\
				5& 500& 	0.0025 	$\pm$	0.0008 	&	1.564	$\pm$	5.7618 	&	0.0155 	$\pm$	0.0891 	&	3 	\\
				&1000&	0.0026 	$\pm$	0.0006 	&	1.028	$\pm$	4.9323 	&	0.0185 	$\pm$	0.1854 	&	1 	\\
				&10000&	0.0149 	$\pm$	0.0058 	&	0.9829	$\pm$	4.7092 	&	0.0231 	$\pm$	0.2308 	&	1 	\\\hline
			\end{tabu}
			\label{tab5}
		\end{table}

	\section{Concluding remarks}
	This paper established a new PE  condition for uniquely determining the parameters of SL systems by the data, which is tighter than existing PE conditions given in the literature.
	A block coordinate descent algorithm is used to solve the optimization problem for identifying the parameters and membership indices in a heuristic way and a theoretical algorithm for  asymptotically finding the number of subsystems is proposed when it is unknown.

	\bibliographystyle{plain}
	\bibliography{ref}
	
\end{document}